\documentclass[referee]{raa}            
\usepackage{graphicx,times}             
\usepackage{gensymb}
\usepackage{float}

\begin{document}

   \title{Study of the long-term $BVR_{c}I_{c}$ photometric variability of eight PMS stars in the young open cluster Trumpler 37 $^*$
\footnotetext{$*$ The photometric data will be available in electronic form via CDS VizieR Online Data Catalogue.}
}

   \volnopage{Vol.0 (20xx) No.0, 000--000}      
   \setcounter{page}{1}          

   \author{Sunay Ibryamov
      \inst{1}
   \and Gabriela Zidarova
      \inst{1}
   \and Evgeni Semkov
      \inst{2}
   \and Stoyanka Peneva
      \inst{2}
   }

   \institute{Department of Physics and Astronomy, Faculty of Natural Sciences, University of Shumen,
             115, Universitetska Str., 9700 Shumen, Bulgaria; {\it sibryamov@shu.bg}\\
        \and
             Institute of Astronomy and National Astronomical Observatory, Bulgarian Academy of Sciences, 72, Tsarigradsko Shose Blvd., 1784 Sofia, Bulgaria\\
   }

   \date{Received~~...; accepted~~...}

\abstract{This paper reports results from our long-term $BV(RI)_{c}$ photometric CCD observations of eight pre-main-sequence stars collected from June 2008 to October 2022.
These stars are located in the young open cluster Trumpler 37, in the field of GM Cephei.
The observational data indicate that all stars from our study exhibit variability in all-optical passbands, typical for young stars.
In this paper, we describe and discuss the photometric behavior of the stars and the possible reasons for their variability.
For two of the objects, we identified periodicity in their light variation.
}

   \authorrunning{S. Ibryamov, G. Zidarova, E. Semkov, S. Peneva}            
   \titlerunning{Study of the long-term photometric variability of eight PMS stars in Trumpler 37}  

   \maketitle

%
%
\section{Introduction}           
\label{sect:intro}

Trumpler 37 is a young open cluster and it is immersed in the HII region IC 1396.
This cluster was identified with the nucleus of the Cepheus OB2 association (Simonson~\cite{simo68}).
Patel et al.~(\cite{pate95}) suggested that the stars forming in the globules of IC 1396 represent the youngest generation of stars in the Cepheus region.
Marschall et al. (1990) obtained photoelectric $UBV$ observations of 120 stars in Trumpler 37 and derived its age of 6.7 Myr.
The authors also established that the cluster contains many possible pre-main-sequence (PMS) stars.
The latest age estimation of Trumpler 37 yields $\sim$4 Myr (Sicilia-Aguilar et al.~\cite{sici05}) and the distance determined to it by Contreras et al.~(\cite{cont02}) is 870 pc.
The cluster was an object of many studies, the recently of which were published by Sicilia-Aguilar et al.~(\cite{sici05}), Sicilia-Aguilar et al.~(\cite{sici06}), Sicilia-Aguilar et al.~(\cite{sici10}), Barentsen et al.~(\cite{bare11}), Meng et al.~(\cite{meng19}), etc.

PMS stars are young stars in the early stages of stellar evolution.
They are divided into two main groups, which are T Tauri stars (TTSs) with relatively low mass (M$\leq$2M$_{\odot}$) and the more massive (2M$_{\odot}$$\leq$M$\leq$8M$_{\odot}$) Herbig Ae/Be stars (HAEBEs).
One of the main features of both groups of PMS stars is their photometric variability.

The study of TTSs began in the middle of the 20th century by Joy~(\cite{joy45}).
TTSs are separated into two subgroups $-$ classical T Tauri stars (CTTSs) and weak-line T Tauri stars (WTTSs).
The CTTSs are surrounded by spacious accreting circumstellar disks, whereas they must have almost disappeared (at least their inner parts) in the WTTSs (M\'{e}nard \& Bertout~\cite{mena99}).
The presence or absence of an active accretion disk determines most of the features that characterize each subgroup of TTSs.
Reviews of the characteristics of TTSs were given in the works of Cram et al.~(\cite{cram89}) and Petrov~(\cite{petr03}).

A detailed study of the various causes of the variability of TTSs was carried out by Herbst et al.~(\cite{herb94}), based on a large electronic $UBVRI$ catalogue that includes several hundred stars.
According to Herbst et al.~(\cite{herb94}), the variability of WTTSs is due to cool spots on the stellar surface.
The periods of this variability are observed on timescales of days and with amplitudes up to 0.8 mag ($V$) and 0.5 mag ($I$).
WTTSs can also show flares related to surface magnetic activity.
The causes of the variability of CTTSs is more complicated.
These stars exhibit irregular brightness variations on timescales of hours and with amplitudes up to 2.6 mag ($V$) and 1.6 mag ($I$).
The variations in mass accretion rate and the presence of hot and cool (in some CTTSs) spots on the stellar surface are responsible for the variability of the CTTSs.

In some young stars large amplitude dips in their brightness can be observed.
These objects are called UXors, named after their prototype UX Orionis.
The UXors generally exhibit irregular variations on timescales of days to weeks and with amplitudes up to 2.8 mag.
It is generally accepted that the registered minima in the brightness of these objects result from the variable circumstellar obscuration (Zaitseva~\cite{zait86}, Voshchinnikov~\cite{vosh89}, Grinin et al.~\cite{grin91}, Herbst et al.~\cite{herb94}).
During the deep minima, the UXors often become bluer as they fade.
This is the so-called \textit{blueing effect} or \textit{color reverse} (see Bibo \& Th\'{e}~\cite{bibo90}).
The reason for this effect is that when the circumstellar clouds of dust and gas or cometary bodies, orbiting the central star, cross the line of sight, a decrease in the brightness of the star occurs.
Initially, the star becomes redder, but during a large extinction the scattered light from the dust clumps begins to dominate and the star becomes bluer.
In the work of Dullemond et al.~(\cite{dull03}) the UXor' variability was explained with self-shadowed disks.

In this study, we present results from the $BV(RI)_{c}$ photometric monitoring of eight stars, located in the cluster Trumpler 37, in the field of GM Cep.
On the basis of the long-term light curves of the stars, we discuss their photometric behavior and the probably causes of their variability.
Our studies of the star GM Cep itself were published in Semkov \& Peneva~(\cite{semk12}), Semkov et al.~(\cite{semk15}) and Mutafov et al.~(\cite{muta22}).
In Section 2 of this paper, we give information about the telescopes and CCD cameras used and the process of data reduction.
In Section 3 we describe the results obtained for the investigated stars and their interpretation.
Finally, in Section 4 we briefly present the concluding remarks.

\section{Observations and data reduction}
\label{sect:Obs}

Our observations of the field centered on GM Cep were obtained during the period from June 2008 to October 2022 with four telescopes and six different CCD cameras, as follows $-$ the 2-m Ritchey-Chr\'{e}tien-Coud\'{e} (with cameras VersArray 1300B and Andor iKon-L), the 50/70-cm Schmidt (with cameras SBIG STL-11000M and FLI PL16803) and the 60-cm Cassegrain (with camera FLI PL09000) telescopes of the Rozhen National Astronomical Observatory in Bulgaria, and the 1.3-m Ritchey-Chr\'{e}tien (with camera Andor DZ436-BV) telescope of the Skinakas Observatory\footnote{Skinakas Observatory is a collaborative project of the University of Crete, the Foundation for Research and Technology, Greece, and the Max-Planck-Institut f{\"u}r Extraterrestrische Physik, Germany.} of the University of Crete in Greece.
The observational procedure and the technical parameters of the CCD cameras used are given in Ibryamov et al.~(\cite{ibry15}) and Ibryamov \& Semkov~(\cite{ibry20}).

All frames were obtained through a standard Johnson$-$Cousins (\textit{BVR$_{c}$I$_{c}$}) set of filters.
The observational data were reduced by applying standard \textsc{idl} procedures, adapted from \textsc{daophot}.
As a reference, the comparison sequence reported in Semkov \& Peneva~(\cite{semk12}) was used.
All data were analyzed using the same aperture, which was chosen to have a 6 arcsec radius, while the background annulus was taken from 10 arcsec to 15 arcsec.
The mean value of errors in the reported magnitudes is 0.01-0.03 mag for the $I_{c}$ and $R_{c}$ bands data, 0.01-0.04 mag for the $V$ band data, and 0.01-0.05 mag for the $B$ band data.

\section{Results and discussion}
\label{sect:Res}

\begin{figure*}[h!]
	\centering
	\includegraphics[width=\textwidth]{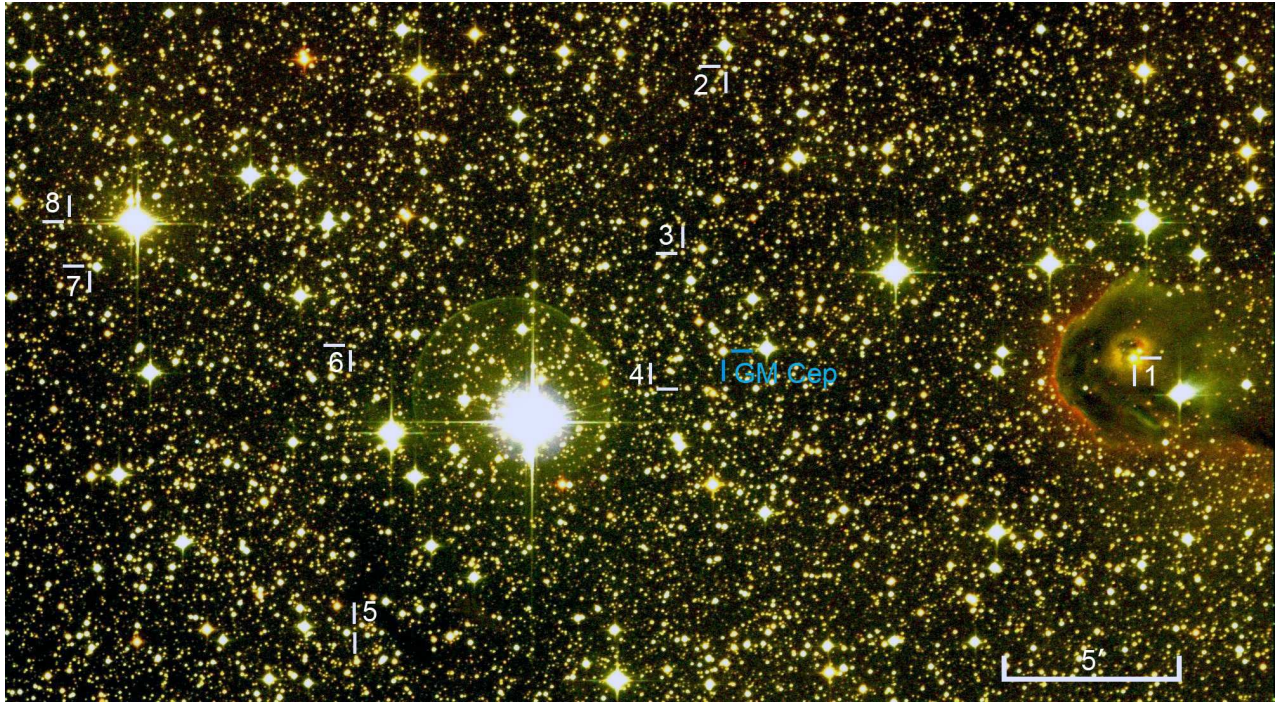}
	\caption{A three-color crop image of the field, where the positions of the stars from our study (designated using their number from Table~\ref{Tab:designations}) and GM Cep are marked. North is up, and East is left. The image was obtained with the 50/70-cm Schmidt telescope.}\label{Fig:field}
\end{figure*}

We listed the stars from our study in Table~\ref{Tab:designations}.
The table columns contain the 2MASS designations, equatorial coordinates, and the number of photometric points obtained for each star with different telescopes used.
Because of the different fields of view of the telescopes and cameras used, we have a different number of photometric points for each star.
Figure~\ref{Fig:field} shows a three-color image of the field, where the positions of the stars from our study and GM Cep are marked.

{\footnotesize
\begin{table*}[h!]
  \caption{Designations, equatorial coordinates and number of observations of the stars.}\label{Tab:designations}
  \begin{center}
  \begin{tabular}{cccccccc}
	  \hline\hline
	  \noalign{\smallskip}
Nr. & 2MASS ID          & RA$_{J2000.0}$ & Dec$_{J2000.0}$ & Nr. of obs. & Nr. of obs.   & Nr. of obs.     & Nr. of obs.   \\
    &                   &                    &                     & (2-m tel.) & (1.3-m tel.) & (Schmidt tel.) & (60-cm tel.) \\            
1   & J21365072+5731106 & 21:36:50.72        & +57:31:10.7         & -          & -            & 208            & 6            \\
2   & J21381703+5739265 & 21:38:17.03        & +57:39:26.6         & -          & -            & 222            & 6            \\
3   & J21382596+5734093 & 21:38:25.97        & +57:34:09.4         & 66         & 51           & 224            & 11           \\
4   & J21383255+5730161 & 21:38:32.55        & +57:30:16.1         & 82         & 51           & 220            & 11            \\
5   & J21393480+5723277 & 21:39:34.81        & +57:23:27.8         & -          & -            & 217            & 6             \\
6   & J21393612+5731289 & 21:39:36.13        & +57:31:28.9         & -          & -            & 221            & 5            \\
7   & J21403134+5733417 & 21:40:31.35        & +57:33:41.8         & -          & -            & 163            & -             \\
8   & J21403574+5734550 & 21:40:35.75        & +57:34:55.1         & -          & -            & 145            & -             \\
   	  \hline \hline
  \end{tabular}
  \end{center}
	\end{table*}}

Table~\ref{Tab:photometry} presents a small fraction of the photometric CCD observations of 2MASS J21382596+5734093 as a guideline on the form and content of the full tables for each star.
The table columns contain Date (dd.mm.yyyy format) and Julian date (JD) of the observation, the measured magnitudes of the star, and the telescope and CCD camera employed.

{{\footnotesize
	\begin{table*}[h!]
		\caption{Photometric CCD observations of 2MASS J21365072+5731106.}\label{Tab:photometry}
		\begin{center}
			\begin{tabular}{cccccccc}
				\hline
				\noalign{\smallskip}
Date       & JD (24...)   & $I_{c}$ & $R_{c}$ & $V$   & $B$   & Tel   & CCD \\
\hline 
28.06.2008 &	54646.424 &	14.56 &	15.65 & 16.68 &	18.106 & 1.3-m RC &	Andor \\
29.06.2008 &	54647.433 &	14.53 &	15.63 &	16.64 &	18.098 & 1.3-m RC &	Andor \\
05.07.2008 &	54653.367 &	-     &	15.46 &	16.43 &	17.868 & 1.3-m RC &	Andor \\
06.07.2008 &	54654.428 &	14.49 &	15.65 &	16.66 &	18.119 & 1.3-m RC &	Andor \\
08.07.2008 &	54656.463 &	14.50 &	15.62 &	16.59 &	18.059 & 1.3-m RC &	Andor \\
13.07.2008 &	54661.428 &	14.40 &	15.65 &	16.39 &	17.808 & 1.3-m RC &	Andor \\
24.07.2008 &	54672.334 &	14.44 &	15.49 &	16.48 &	17.992 & 1.3-m RC &	Andor \\
25.07.2008 &	54673.354 &	14.55 &	15.63 &	16.61 &	18.055 & 1.3-m RC &	Andor \\
02.08.2008 &	54680.530 &	14.46 &	15.56 &	16.53 &	18.036 & 1.3-m RC &	Andor \\
27.08.2008 &	54706.291 &	14.44 &	15.46 &	16.41 &	-      & Schmidt  &	STL-11 \\
...        &    ...       & ...     & ...     & ...   & ...   &   ... & ... \\
				\hline
			\end{tabular}
		\end{center}
{\textbf{Note:} The full tables will be available in electronic form via CDS VizieR Online Data Catalogue}.
\end{table*}}

In Table~\ref{Tab:amplitudes} we report the $BV(RI)_{c}$ maximal and minimal magnitudes and the amplitudes of the variability of the investigated stars registered during our monitoring.
It can be seen from Table~\ref{Tab:amplitudes} that the lowest amplitude star is 2MASS J21365072+5731106 ($\Delta{V}$=0.14 mag), the $V$ amplitude of six stars is in the range from 0.56 to 1.54 and the largest amplitude star is 2MASS J21393480+5723277 ($\Delta{V}$=2.03 mag).
The stellar parameters of the objects from the literature are summarized in Table~\ref{Tab:parameters}.

\begin{table}[h!]
	{\footnotesize
		\caption{The registered $BV(RI)_{c}$ maximal and minimal magnitudes and the amplitudes of the variability of the stars.}\label{Tab:amplitudes}
		\begin{center}
			\begin{tabular}{lccccccccccccc}
				\hline\hline
				\noalign{\smallskip}
Nr. & Star (2MASS ID)   & $B_{max}$ & $B_{min}$ & $V_{max}$ & $V_{min}$ & $R_{c max}$ & $R_{c min}$ & $I_{c max}$ & $I_{c min}$ & $\Delta{B}$ & $\Delta{V}$ & $\Delta{R_{c}}$ & $\Delta{I_{c}}$ \\ 		
				\noalign{\smallskip}
				\hline
				\noalign{\smallskip}
1  & J21365072+5731106 & 15.15 & 15.33 & 13.50 & 13.64 & 12.43 & 12.58 & 11.27 & 11.41 & 0.18 & 0.14 & 0.15 & 0.14 \\
2  & J21381703+5739265 & 18.13 & 19.38 & 16.54 & 17.78 & 15.53 & 16.66 & 14.40 & 15.25 & 1.25 & 1.24 & 1.13 & 0.85 \\
3  & J21382596+5734093 & 17.60 & 18.92 & 16.24 & 17.24 & 15.21 & 16.18 & 14.27 & 15.00 & 1.32 & 1.00 & 0.97 & 0.73 \\
4  & J21383255+5730161 & 17.79 & 19.15 & 16.55 & 17.63 & 15.73 & 16.82 & 14.68 & 16.23 & 1.36 & 1.08 & 1.09 & 1.55 \\
5  & J21393480+5723277 & 18.42 & 19.49 & 17.18 & 19.21 & 16.24 & 17.85 & 15.15 & 16.70 & 1.07 & 2.03 & 1.61 & 1.55 \\
6  & J21393612+5731289 & 17.80 & 18.54 & 16.49 & 17.05 & 15.70 & 16.27 & 14.64 & 15.09 & 0.74 & 0.56 & 0.57 & 0.45 \\
7  & J21403134+5733417 & 16.90 & 18.41 & 15.83 & 17.37 & 15.28 & 16.81 & 14.38 & 15.79 & 1.51 & 1.54 & 1.53 & 1.41 \\
8  & J21403574+5734550 & 17.74 & 18.74 & 16.57 & 17.29 & 15.68 & 16.26 & 14.74 & 15.20 & 1.00 & 0.72 & 0.58 & 0.46 \\
				\hline \hline
			\end{tabular}
	\end{center}}
\end{table}

{\footnotesize
	\begin{table*}[h!]
		\caption{Stellar parameters of the stars.}\label{Tab:parameters}
		\begin{center}
			\begin{tabular}{ccccccc}
				\hline\hline
				\noalign{\smallskip}
Nr. & Star (2MASS ID)   & Spectral type & Luminosity (L$_{\odot}$) & Radius (R$_{\odot}$) & Mass (M$_{\odot}$) & Reference \\
\hline
1 & J21365072+5731106 & F8             & 84                       & 8.4                  & $>$3.0              & 1          \\
2 & J21381703+5739265 & K7            & 1.16                     & 2.2                  & 0.8                  & 2         \\
3 & J21382596+5734093 & K5.5          & 1.12                     & 2.0                  & 1.0                  & 2         \\
4 & J21383255+5730161 & K7            & -                        & -                    & 0.8                  & 3         \\
5 & J21393480+5723277 & K5.5          & 0.67                     & 1.5                  & 1.1                  & 2         \\
6 & J21393612+5731289 & K7.5          & -                        & -                    & 0.7                  & 3         \\ 
7 & J21403134+5733417 & K5            & 1.06                     & 1.9                  & 1.2                  & 2         \\
8 & J21403574+5734550 & K6            & 0.92                     & 1.8                  & 1.1                  & 2         \\
				\hline\hline
			\end{tabular}
		\end{center}
		{\textbf{References:} $^1$Hessman et al.~(\cite{hess95}), $^2$Sicilia-Aguilar et al.~(\cite{sici10}), $^3$Sicilia-Aguilar et al.~(\cite{sici06})}.
\end{table*}}

We used the \textsc{period04} (Lenz \& Breger~\cite{lenz05}) and \textsc{persea} (written by G. Maciejewski, based on Schwarzenberg-Czerny~(\cite{schw96}) software packages to search for periodicity in the photometric variability of the objects.
We discuss the obtained photometric results for the stars in the following subsections.

\textit{\subsection{2MASS J21365072+5731106 (also known as LkH$\alpha$ 349, HBC 308 and V390 Cep)}}

LkH$\alpha$ 349 is a central nebulous star of the nebula IC 1396A and it lies at the focus of a large parabolic rim of emission nebulosity in IC 1396 (see Loren et al.~\cite{lore75} and Nakano et al.~\cite{naka89}).
This star was included in the work of Dibai \& Esipov~(\cite{diba68}), where the authors presented its spectrum.
According to the authors, LkH$\alpha$ 349 exhibits a strong continuous spectrum with sharply defined emission in the H$\alpha$ line and they reported the mean measurements from the photoelectric observations (from July 1966) of the star: $B$=15.25 mag and $V$=13.40 mag.
Dibai \& Esipov~(\cite{diba68}) also noted that the star is very red and that probably most of the reddening is due to the absorption of light in the dense globule.
They are unable to determine the intrinsic color of the star (its spectral type) because of the lack of lines.

LkH$\alpha$ 349 was included in the Second catalog of emission-line stars of the Orion population published by Herbig \& Kameswara Rao~(\cite{herb72}).
Cohen \& Kuhi~(\cite{cohe79}) determined its spectrum as F8 and reported that it has a striking P Cygni profile at H$\alpha$.

Hessman et al.~(\cite{hess95}) conducted a detailed study of LkH$\alpha$ 349 and concluded that it is a PMS star of intermediate mass ($\geq$3M$_{\odot}$) on its way to become a Herbig Be star.
Hessman et al.~(\cite{hess95}) presented a historical photographic $B$ light curve of the star for the period from 1899 to $\sim$1990, obtained from the measurements on plates in the different plate archives.
For this period of time the star's $B$ brightness varies in the range from 13 to 16.2 mag.
Based on the historical light curve, Kazarovets \& Samus~(\cite{kaza97}) cataloged LkH$\alpha$ 349 in the 73$^{rd}$ Name-List of Variable Stars with the designation V390 Cep.

Marschall et al.~(\cite{mars90}) measured $V$=13.41 mag, $B-V$=1.71 mag, $V-R$=1.06 mag and $R-I$=1.07 mag for LkH$\alpha$ 349.
Fernandez~(\cite{fern95}) presented 27 $BV(RI_{c})$ photometric points of the star, obtained in the period from 1991 August to 1992 August.
In the presented analysis of these data in Fernandez \& Eiroa~(\cite{fern96}), the authors concluded that the star's brightness is constant during the observations until the last nights in 1992 August, when it decreased 0.1 mag in all bands.
According to their data the mean value of the star's $V$ brightness is 13.37 mag.

The $BV(RI)_{c}$ light curves of LkH$\alpha$ 349 from our observations (the circles) are shown in Figure~\ref{Fig:curves1}.
For this star, we found photometric data in the database of the American Association of Variable Star Observers (AAVSO, \textit{https://www.aavso.org/}) and included them in the figure (the empty diamonds).
It is seen that there is a good match between our data and ones from AAVSO.
The available data of LkH$\alpha$ 349 suggest that it exhibits variability with small amplitudes.
As can be seen from Fig.~\ref{Fig:curves1} and Table~\ref{Tab:amplitudes}, the registered amplitude of the star's light variation is under 0.2 mag in all-optical passbands.
Variability with such small amplitudes is typical for both weak-line TTSs and HAEBESs, and it likely is produced by cool spots on the star's surface.

\begin{figure}[h!]
	\begin{center}
		\includegraphics[width=15cm]{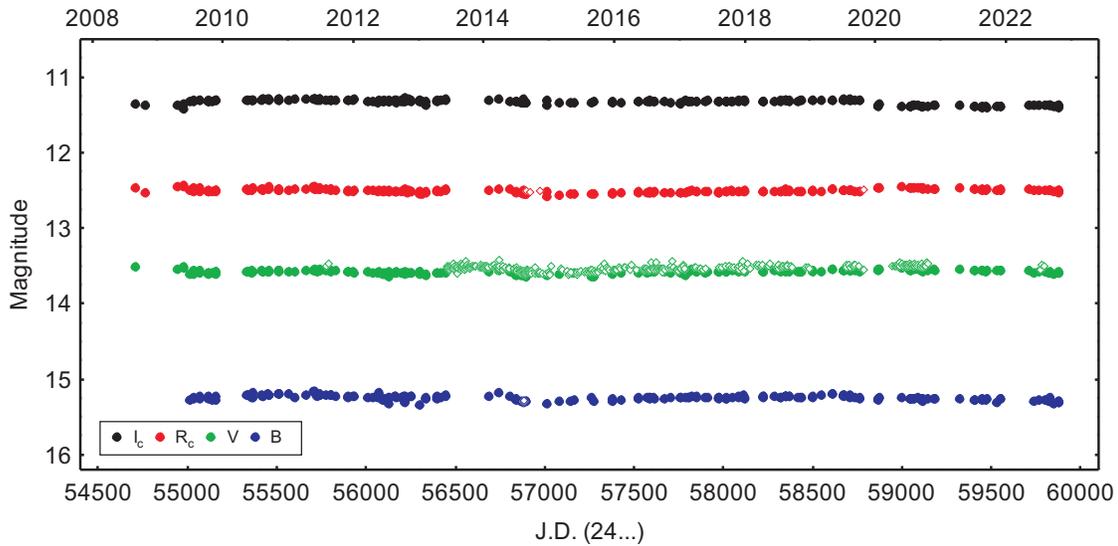}
		\caption{$BVR_{c}I_{c}$ light curves of 2MASS J21365072+5731106. The circles represent our data and the empty diamonds signify the data from AAVSO.}\label{Fig:curves1}
	\end{center}
\end{figure}

The mean values for the brightness of LkH$\alpha$ 349 during the whole time of our monitoring are: $\overline{B}$=15.24 mag, $\overline{V}$=13.57 mag, $\overline{R_{c}}$=12.49 mag and $\overline{I_{c}}$=11.32 mag, and the mean values for its colors are: $\overline{B-V}$=1.67 mag, $\overline{V-R}$=1.08 mag, $\overline{V-I}$=2.25 mag and $\overline{R-I}$=1.18 mag.
By comparing our mean values with those of Dibai \& Esipov~(\cite{diba68}), Marschall et al.~(\cite{mars90}) and Fernandes~(\cite{fern95}) we get very small differences, which are probably due to the different comparison stars used.
Therefore, we can assume that for more than 50 years, the brightness of LkH$\alpha$ 349 maintains its level, and the star's variability is characterized by low amplitude in all-optical passbands.

Using our data and ones from AAVSO, we carried out a periodicity search in the photometric behavior of LkH$\alpha$ 349.
Initially, we used all data obtained during the whole time of the observations, and then the data only from different periods.
However, we did not detect any periodicity in the star's variability.

\textit{\subsection{2MASS J21381703+5739265 and 2MASS J21403574+5734550}}

The light curves of 2MASS J21381703+5739265 and 2MASS J21403574+5734550 from our $BV(RI)_{c}$ monitoring are displayed in Figure~\ref{Fig:curves2} and Figure~\ref{Fig:curves8}, respectively.
The data indicate that the photometric behavior of both stars is characterized by strong irregular variability in all-optical passbands.

It is seen from Fig.~\ref{Fig:curves2} that the total brightness of 2MASS J21381703+5739265 gradually decreases until mid-2014, then it begins to increase and reaches its previous high level towards the end of 2018.
Then, the star's total brightness begins to decrease again.
During our observations, several single dips in the variability of the star are also registered.

The amplitudes of the light variations of 2MASS J21381703+5739265 are given in Table~\ref{Tab:amplitudes}.
Usually, such amplitudes are typical of CTTSs surrounded by a circumstellar disk.
The star's color variations versus the $V$ magnitude (Fig.~\ref{Fig:curves2}) are typical for TTSs.
The possible reasons for the variability of 2MASS J21381703+5739265 can be different, which includes a change in mass accretion rate from the circumstellar disk and the presence of spot(s) on the star's surface.

\begin{figure}[h!]
		\begin{center}
	\includegraphics[width=11cm]{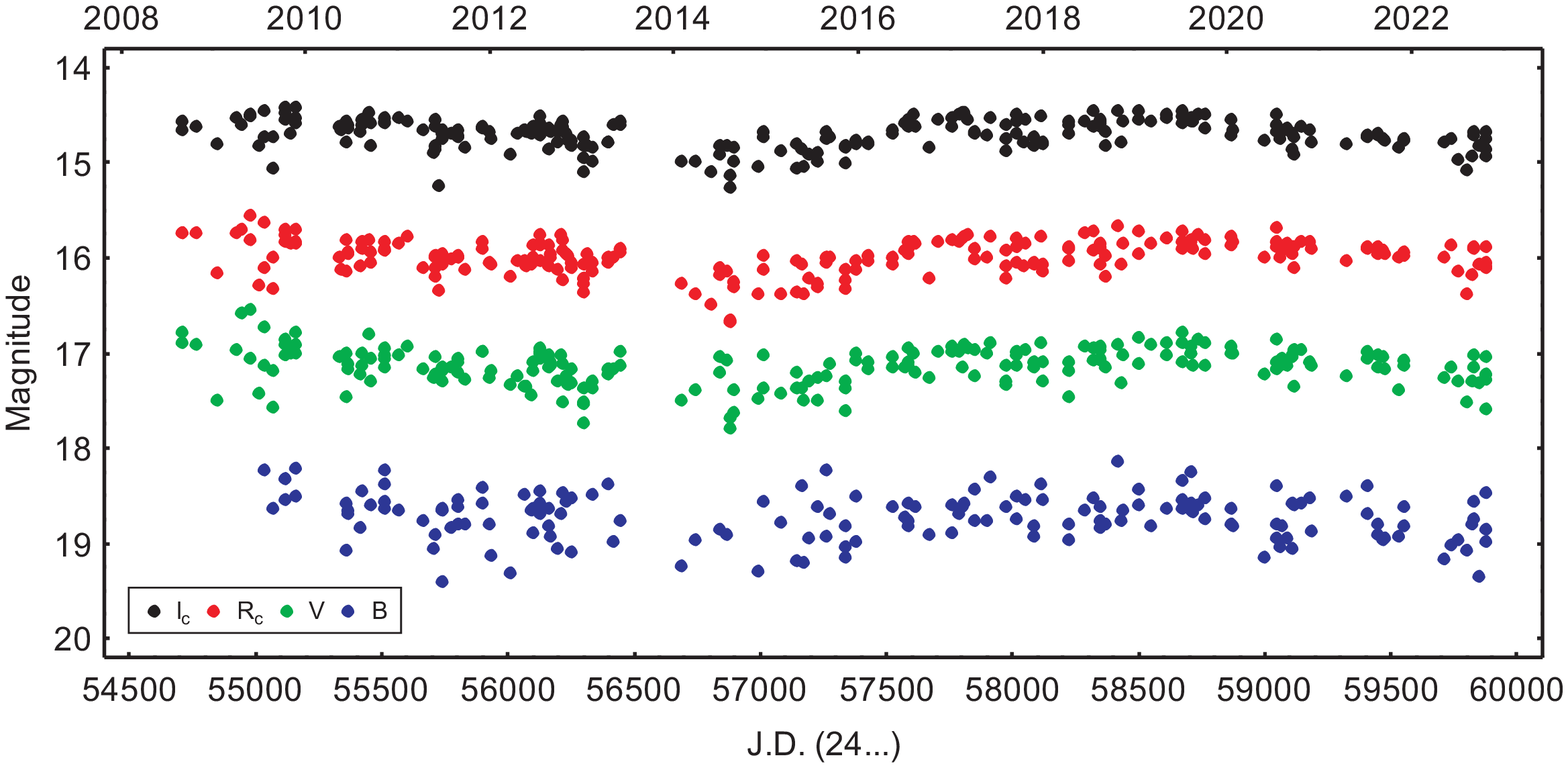}
	\includegraphics[width=3cm]{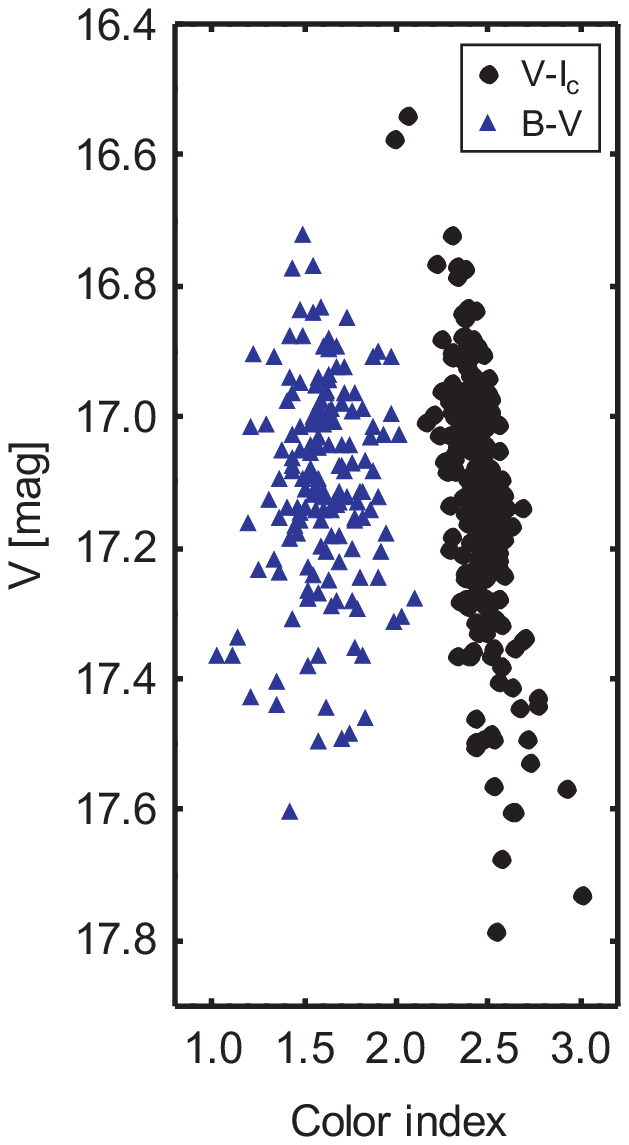}
	\caption{$BVR_{c}I_{c}$ light curves (left) and color$-$magnitude diagrams (right) of 2MASS J21381703+5739265.}\label{Fig:curves2}
		\end{center}
\end{figure}

For 2MASS J21403574+5734550 our data suggest that its brightness generally varies around some intermediate level.
It can be seen from the color$-$magnitude diagrams for the star (Fig.~\ref{Fig:curves8}) that it becomes redder as it fades, typical for TTSs.
The registered amplitudes for the star's variability are given in Table~\ref{Tab:amplitudes}.
It can be assumed that variability with such properties is typical of both subgroups of TTSs and it probably is produced by the rotational modulation of spot(s) on the star's surface.

\begin{figure}[h!]
	\begin{center}
		\includegraphics[width=11cm]{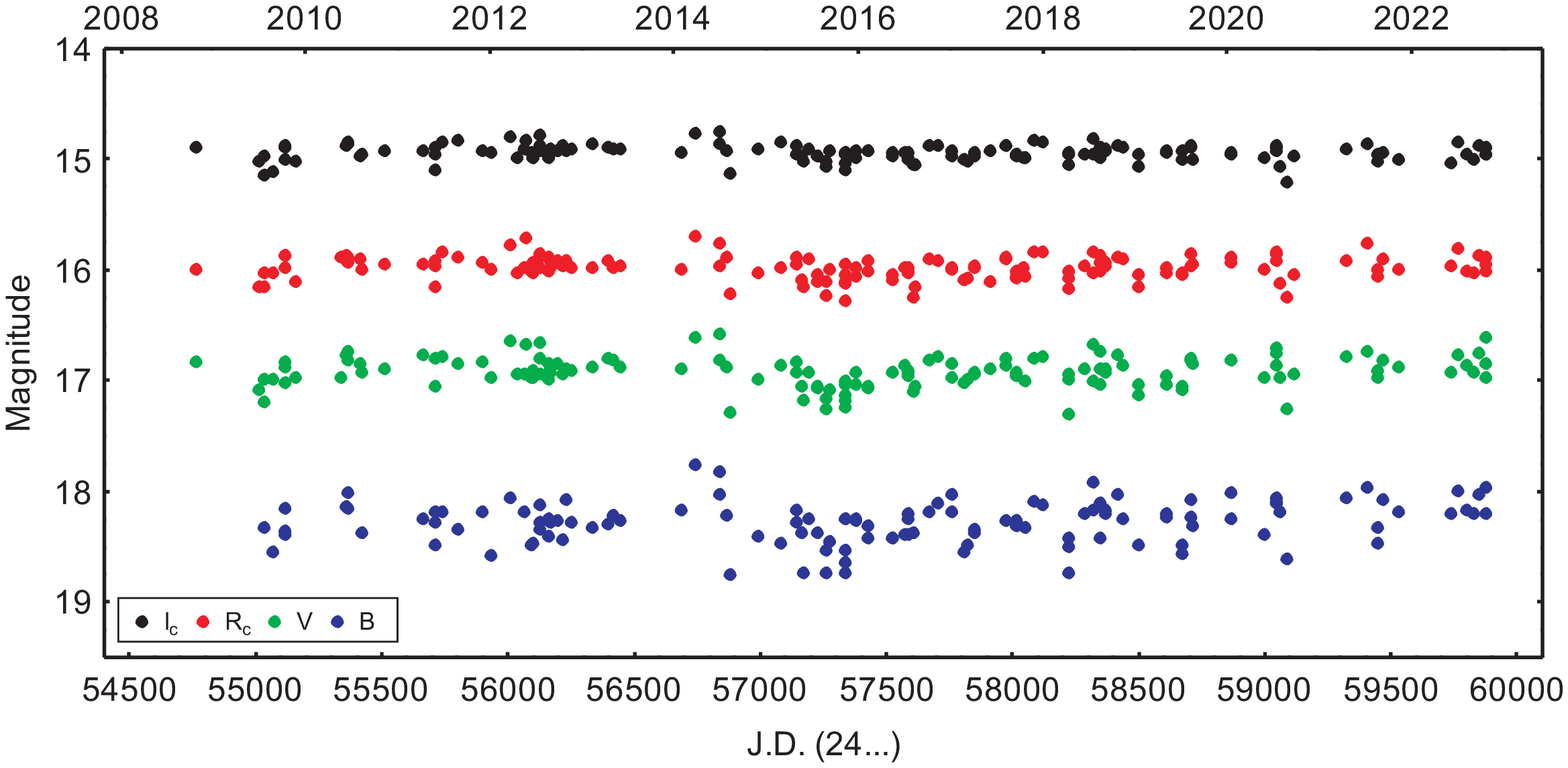}
		\includegraphics[width=3cm]{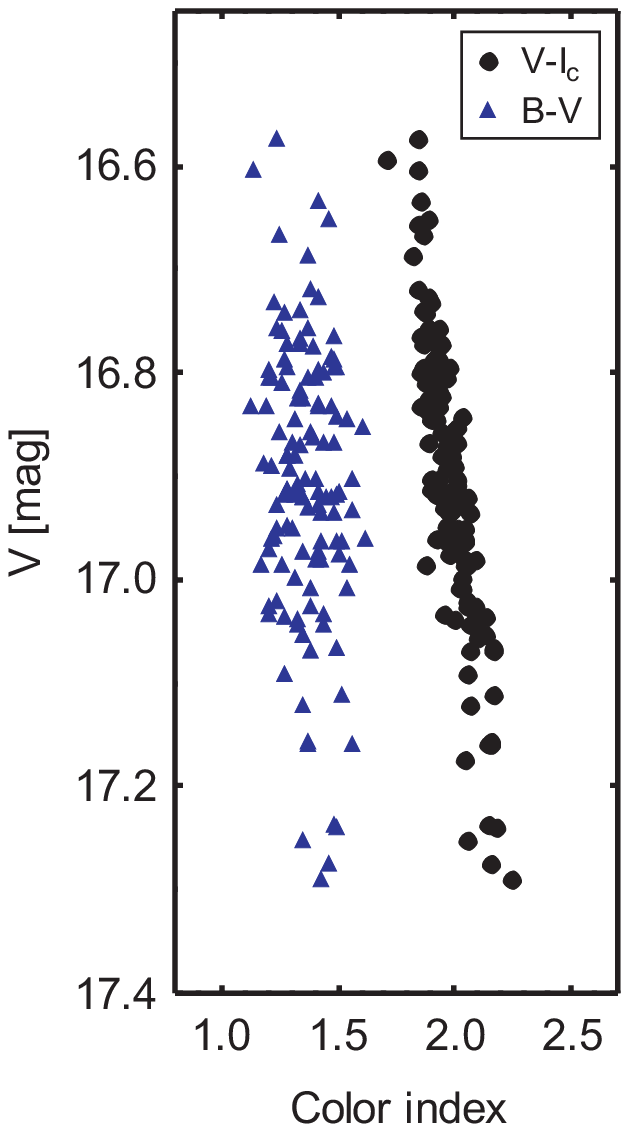}
		\caption{$BVR_{c}I_{c}$ light curves (left) and color$-$magnitude diagrams (right) of 2MASS J21403574+5734550.}\label{Fig:curves8}
	\end{center}
\end{figure}

We did not detect any periodicity in the variability of 2MASS J21381703+5739265 and 2MASS J21403574+5734550.


\textit{\subsection{2MASS J21382596+5734093 and 2MASS J21393612+5731289}}

The $BV(RI)_{c}$ light curves of 2MASS J21382596+5734093 and 2MASS J21393612+5731289 constructed based on our observations are plotted in Figure~\ref{Fig:curves3} and Figure~\ref{Fig:curves6}.

It is seen from Fig.~\ref{Fig:curves3} that from the beginning of our monitoring until 2014 the photometric behavior of 2MASS J21382596+5734093 is characterized by low amplitude variability.
After 2014, the star began to exhibit irregular fading events with different amplitudes in all-optical passbands.
The star's color$-$magnitude diagrams (Fig.~\ref{Fig:curves3}) showed that regardless of the different photometric behavior before and after 2014, during the whole time of our observations, 2MASS J21382596+5734093 becomes redder as it fades.

Our time series analysis of the data for this star gave a positive result.
We identified a rotational period of 0.9030$^d$ in its variability.
This result confirms the star's periodicity given in the Zwicky Transient Facility (ZTF) Catalog of Periodic Variable Stars (Chen et al.~\cite{chen20}).
In Figure~\ref{Fig:period3} are displayed the periodogram and the phased $I_{c}$ and $B$ light curves of the star from our observations.
In the figure, the circles mark the photometric data of the star until 2014, and the empty triangles represent the photometric data after 2014.

\begin{figure}[h!]
	\begin{center}
		\includegraphics[width=11cm]{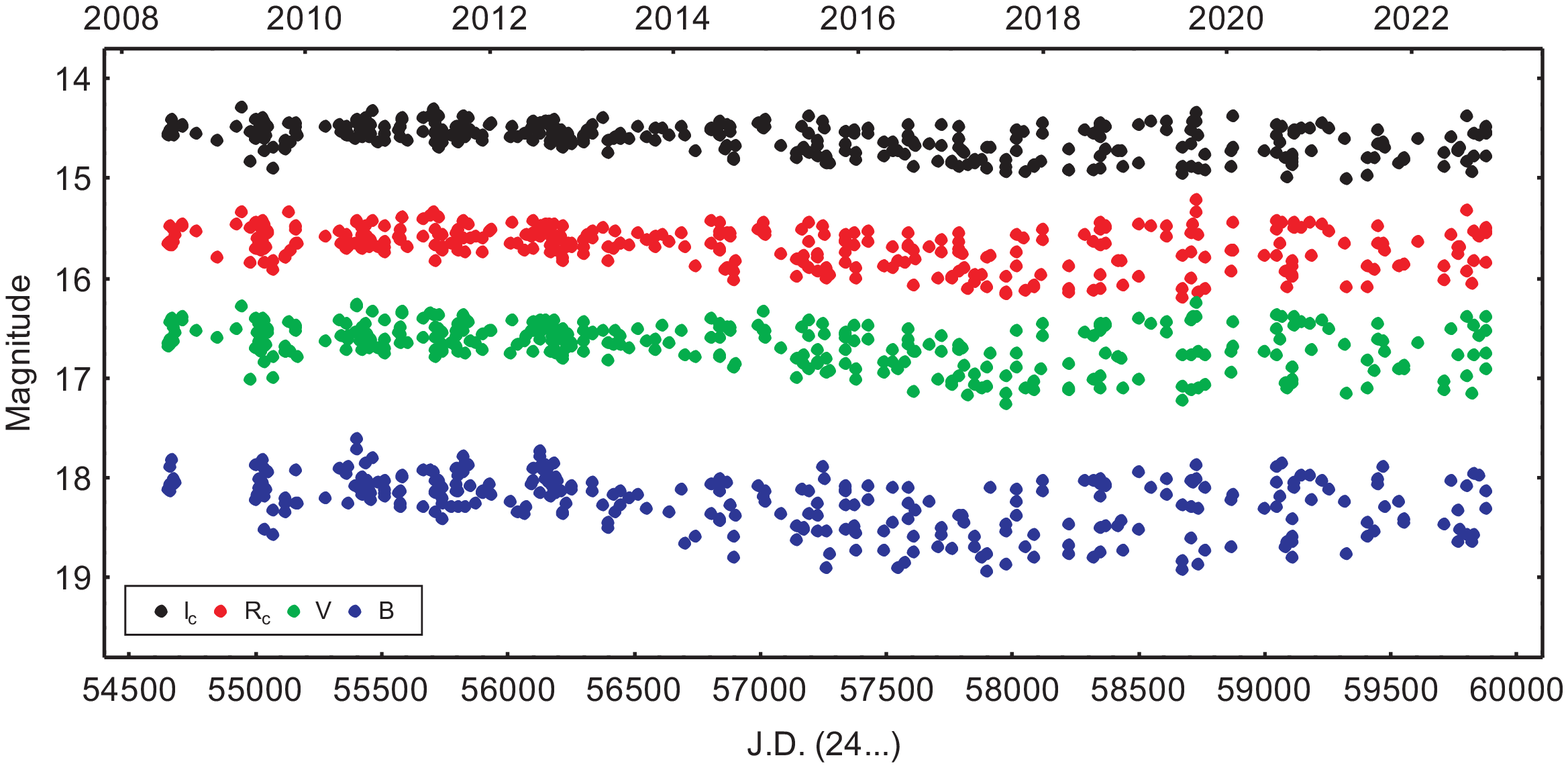}
		\includegraphics[width=3cm]{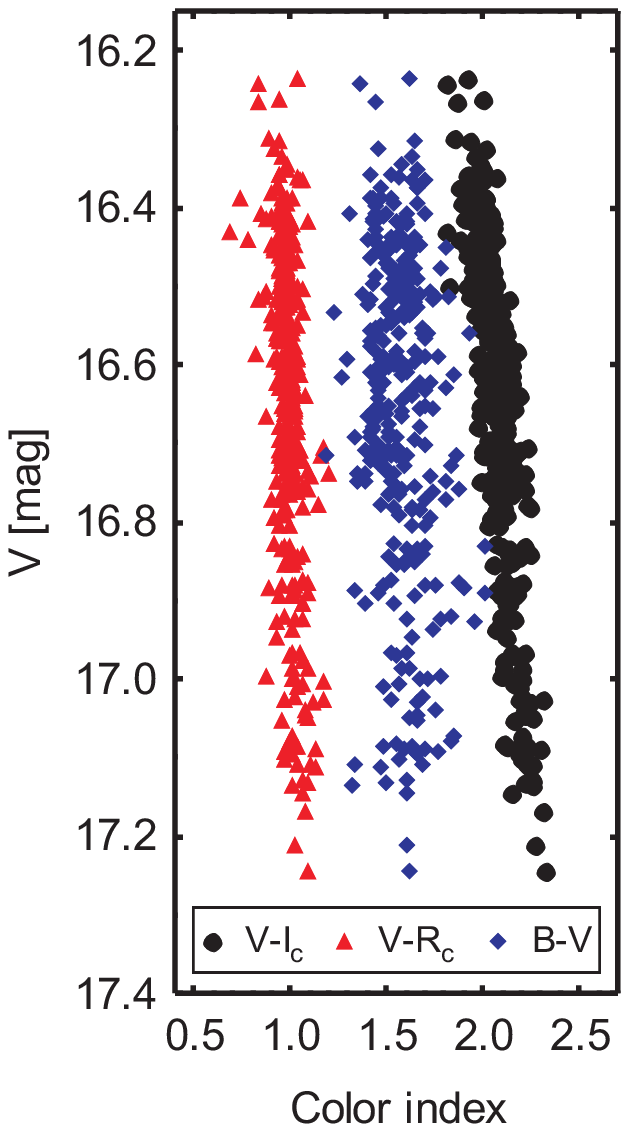}
		\caption{$BVR_{c}I_{c}$ light curves (left) and color$-$magnitude diagrams (right) of 2MASS J21382596+5734093.}\label{Fig:curves3}
	\end{center}
\end{figure}

\begin{figure}[h!]
	\begin{center}
		\includegraphics[width=6.5cm]{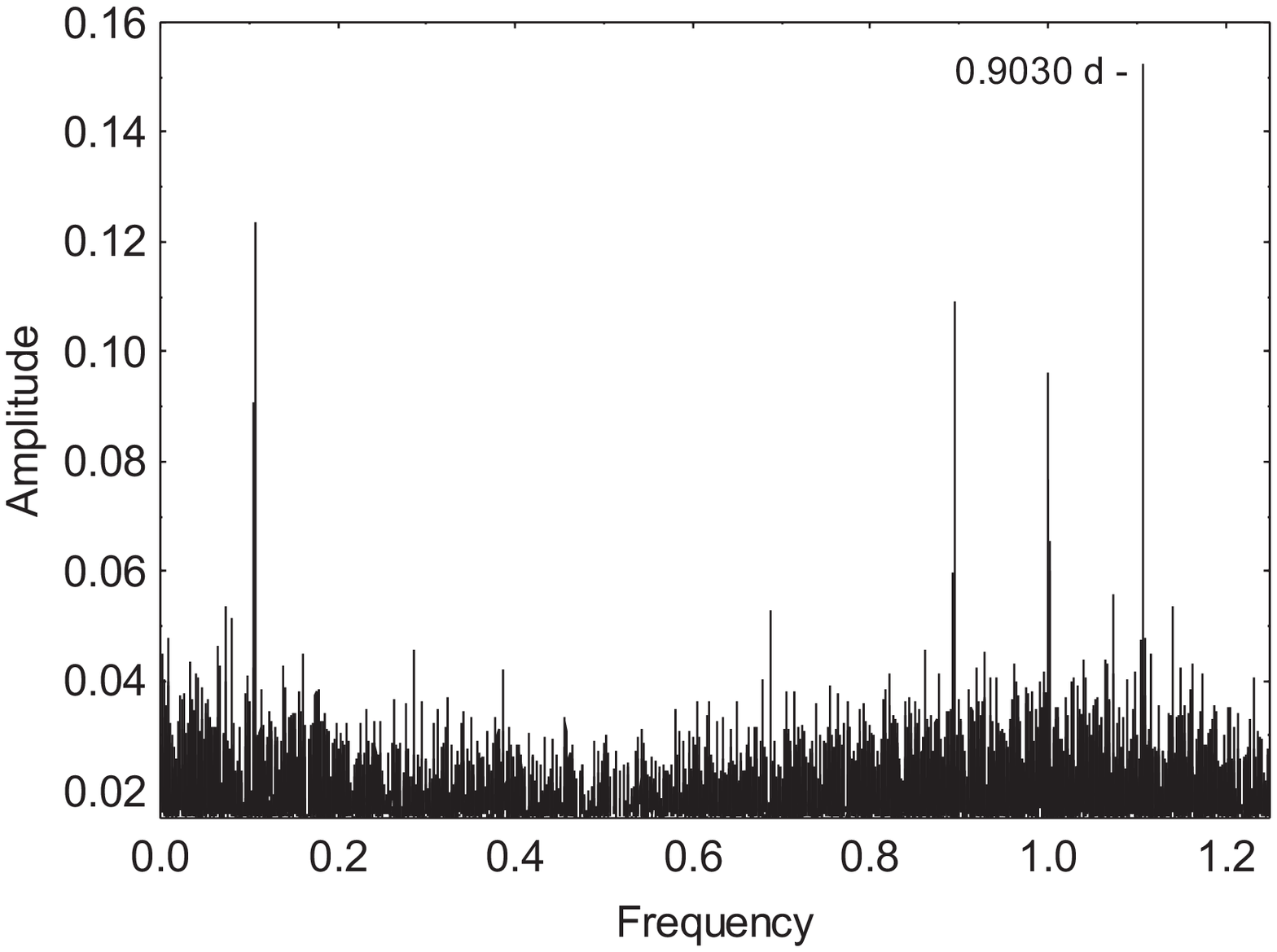}
		\includegraphics[width=6.5cm]{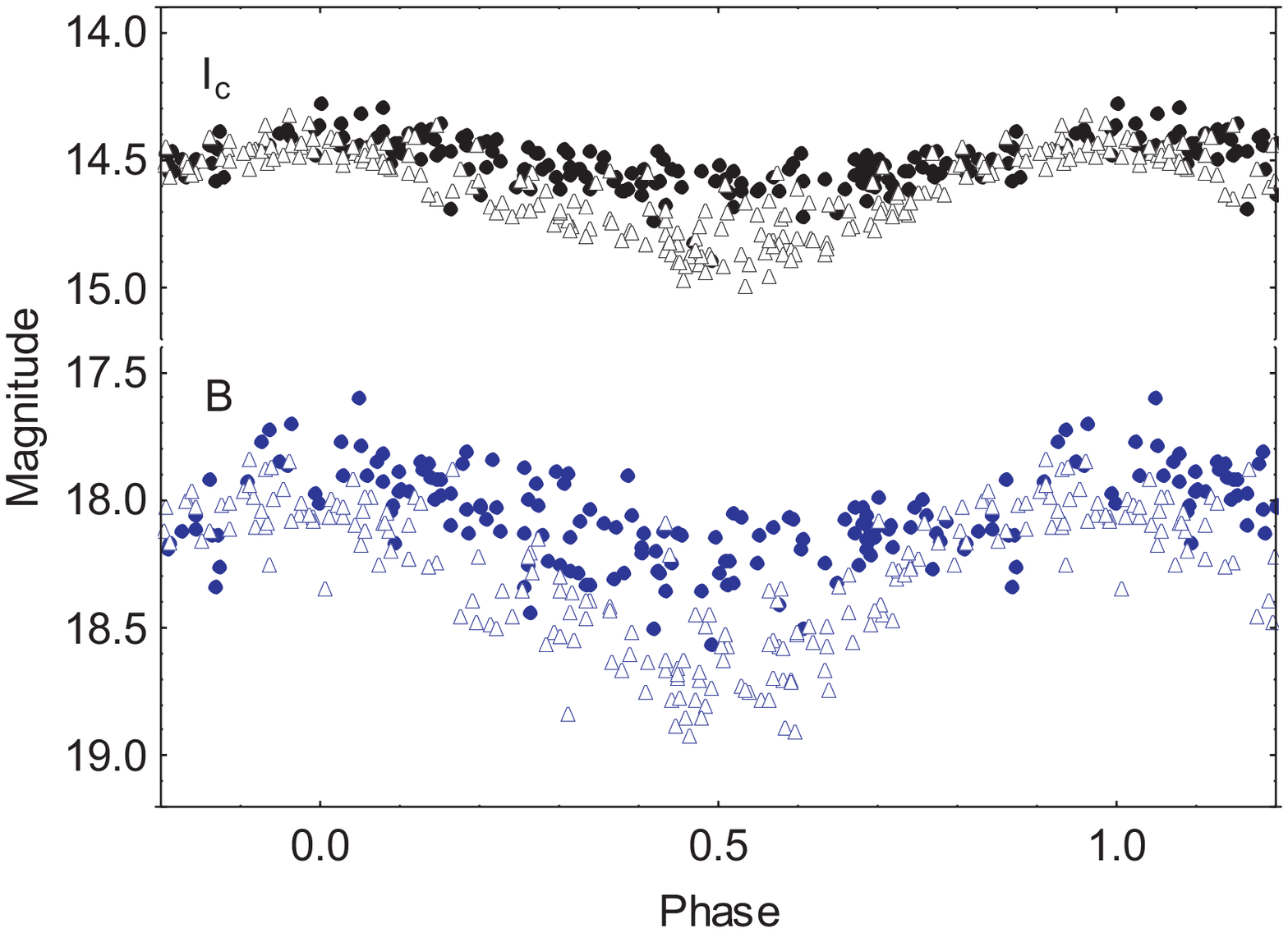}
		\caption{Left: Periodogram of 2MASS J21382596+5734093. Right: $I_{c}$ and $B$ band phased light curves of the star. The circles in the right panel signify star's photometric data until 2014 and the empty triangles mark ones after 2014.}\label{Fig:period3}
	\end{center}
\end{figure}

For 2MASS J21393612+5731289 the available data suggest that it exhibits rapid light variation and its brightness generally varies around some intermediate level in all-optical passbands.
The registered photometric amplitudes of the star's light variations are given in Table~\ref{Tab:amplitudes}.
Our time-series analysis of the data on the star indicated a rotational period of 1.4919$^d$.
The ZTF Catalog of Periodic Variable Stars (Chen et al.~\cite{chen20}) gives a period of 2.9838$^d$ for the star's variability and classified it as an EW-type eclipsing binary star.
As it seen this is twice the period registered by us and this value corresponds to the second peak in our periodogram.
Figure~\ref{Fig:period6} shows the periodogram and the phased $I_{c}$ light curves of the star according to identified both periods.
In the International Variable Star Index (VSX) database of AAVSO, for the periodicity of the star is given the half the ZTF catalogue value, i.e. the value corresponding to the first peak in our periodogram.

\begin{figure}[h!]
	\begin{center}
		\includegraphics[width=14cm]{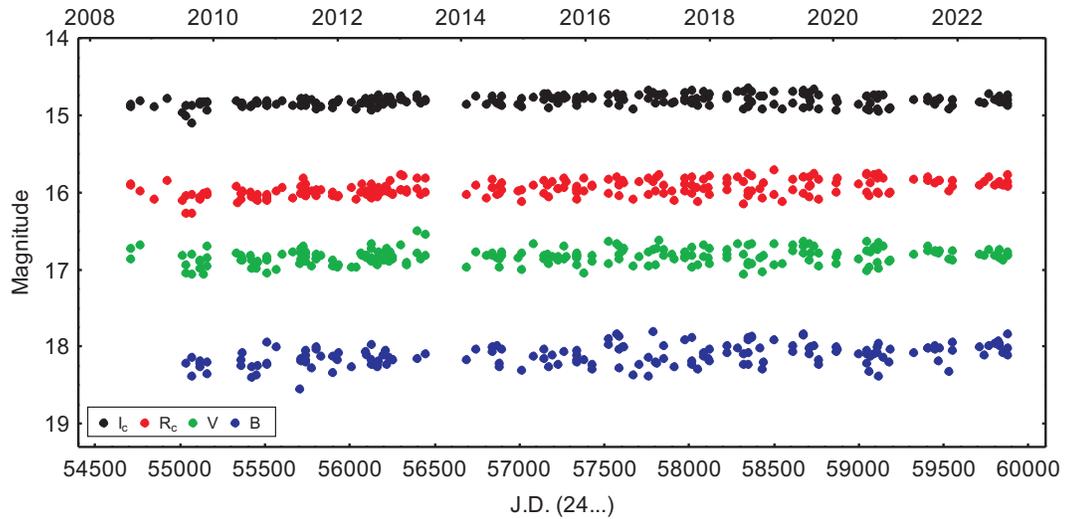}
		\caption{$BVR_{c}I_{c}$ light curves of 2MASS J21393612+5731289.}\label{Fig:curves6}
	\end{center}
\end{figure}

\begin{figure}[h!]
	\begin{center}
		\includegraphics[width=6.5cm]{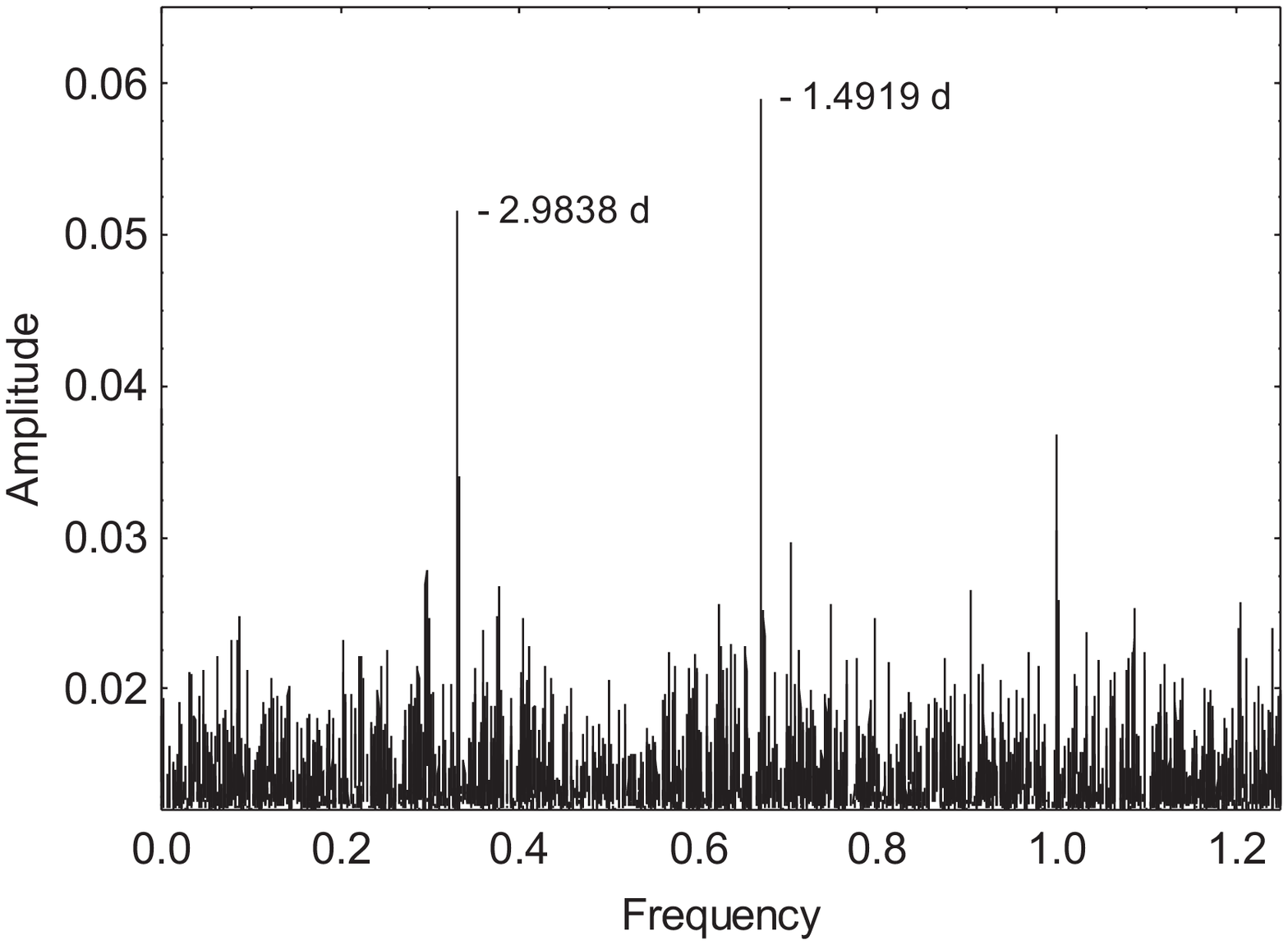}
		\includegraphics[width=7cm]{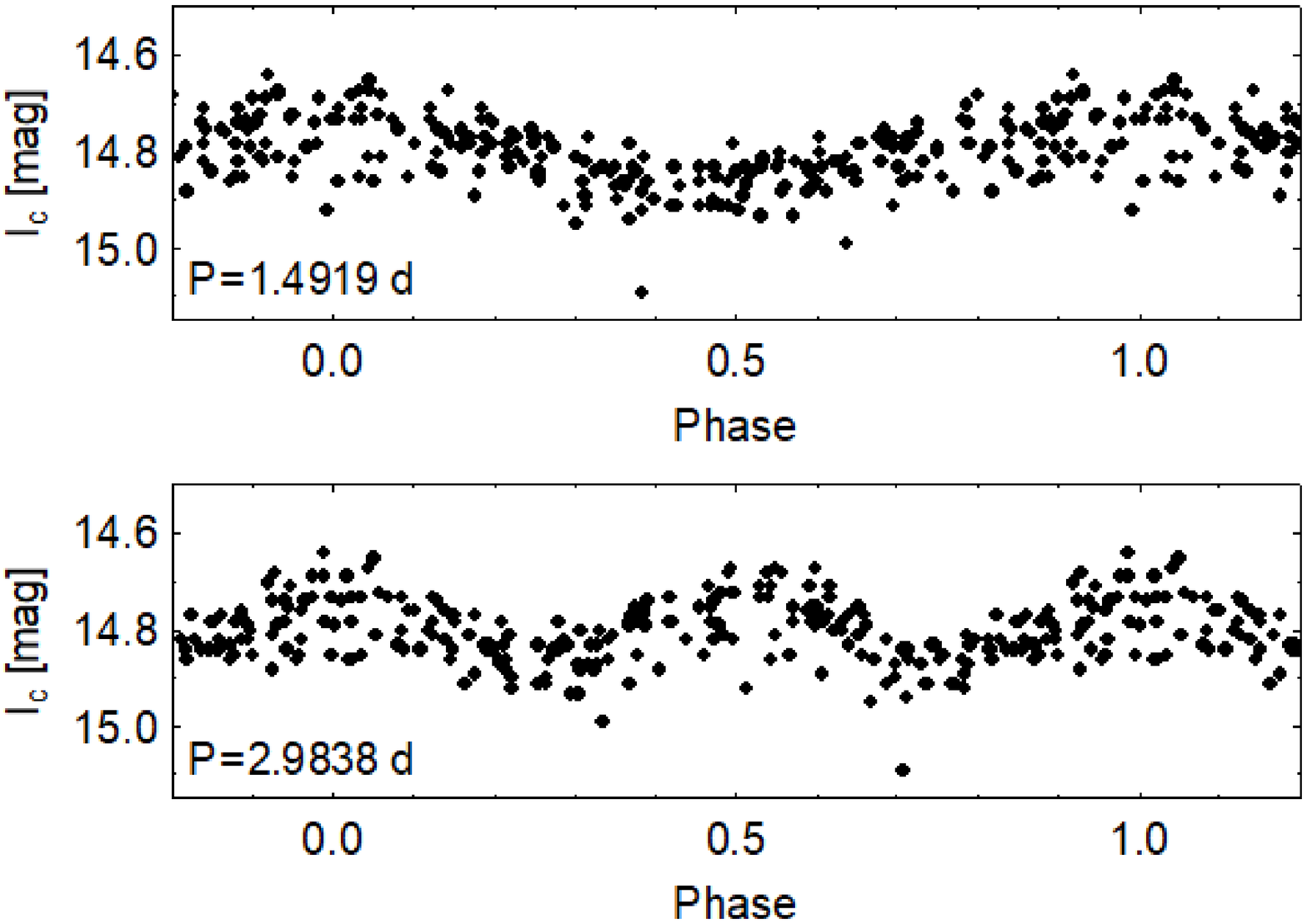}
		\caption{Left: Periodogram of 2MASS J21393612+5731289. Right: $I_{c}$ band phased light curves of the star according to both periods.}\label{Fig:period6}
	\end{center}
\end{figure}

It is important to mention that the registered periodicity in the photometric behavior of 2MASS J21382596+5734093 and 2MASS J21393612+5731289 remained stable during the whole time of our observations (about 18 years), and it is a typical rotational period for a WTTS.
Most probably the periodicity is caused by rotation modulation of cool spot(s) on the stellar surface.

\textit{\subsection{2MASS J21383255+5730161, 2MASS J21393480+5723277 and 2MASS J21403134+5733417}}

The $BV(RI)_{c}$ light curves of 2MASS J21383255+5730161, 2MASS J21393480+5723277 and 2MASS J21403134+5733417 from our monitoring are depicted in Figure~\ref{Fig:curves4}, Figure~\ref{Fig:curves5} and Figure~\ref{Fig:curves7}, respectively.
As can be seen from the figures, the variability of these stars is characterized by multiple fading events with different amplitudes.
For 2MASS J21383255+5730161 these events are distinguishable, but for 2MASS J21393480+5723277 and 2MASS J21403134+5733417 the periods of the dips events are relatively short and sometimes we have only one photometric point in their brightness minima.

\begin{figure}[h!]
	\begin{center}
		\includegraphics[width=11cm]{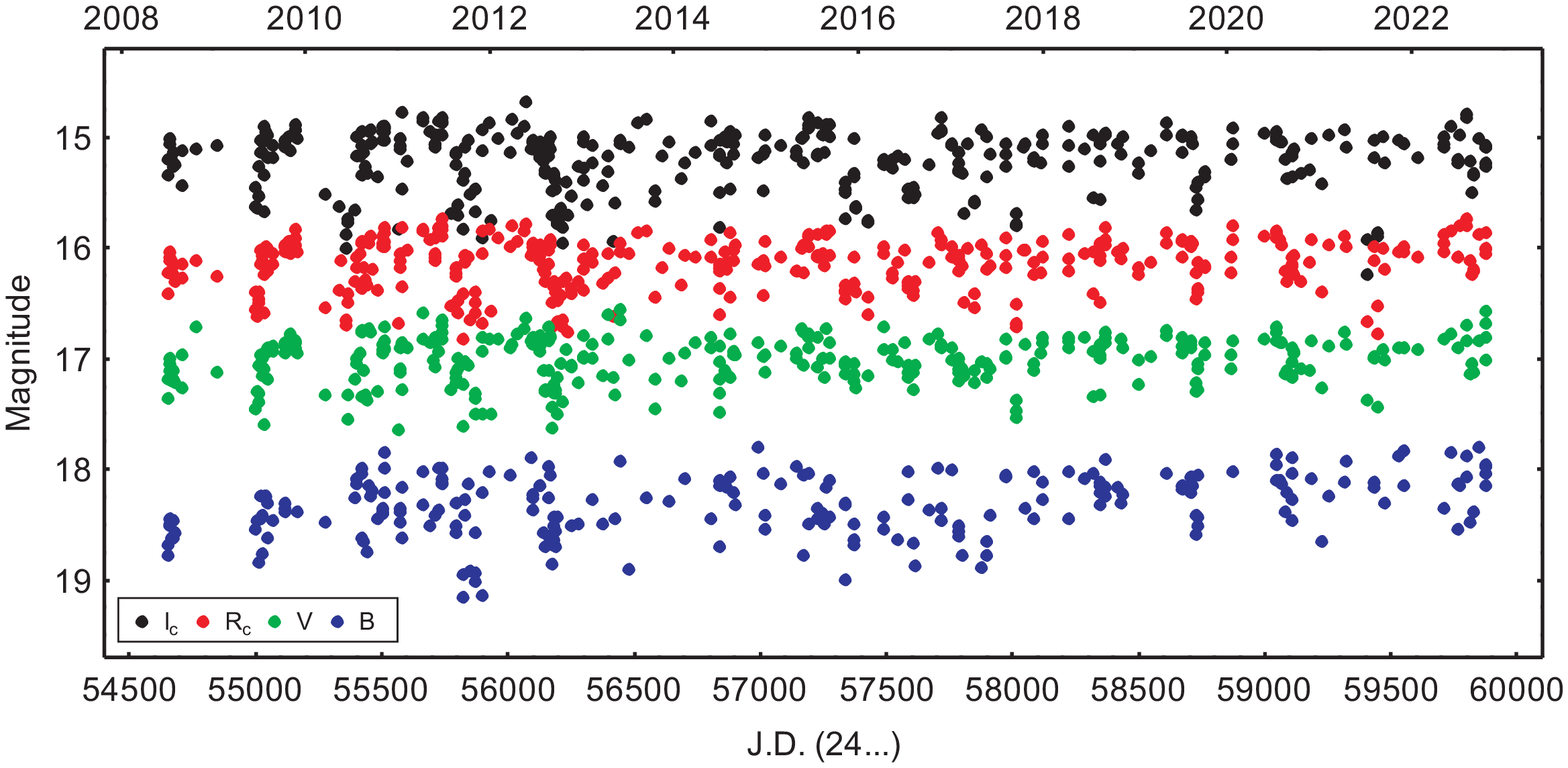}
    	\includegraphics[width=3cm]{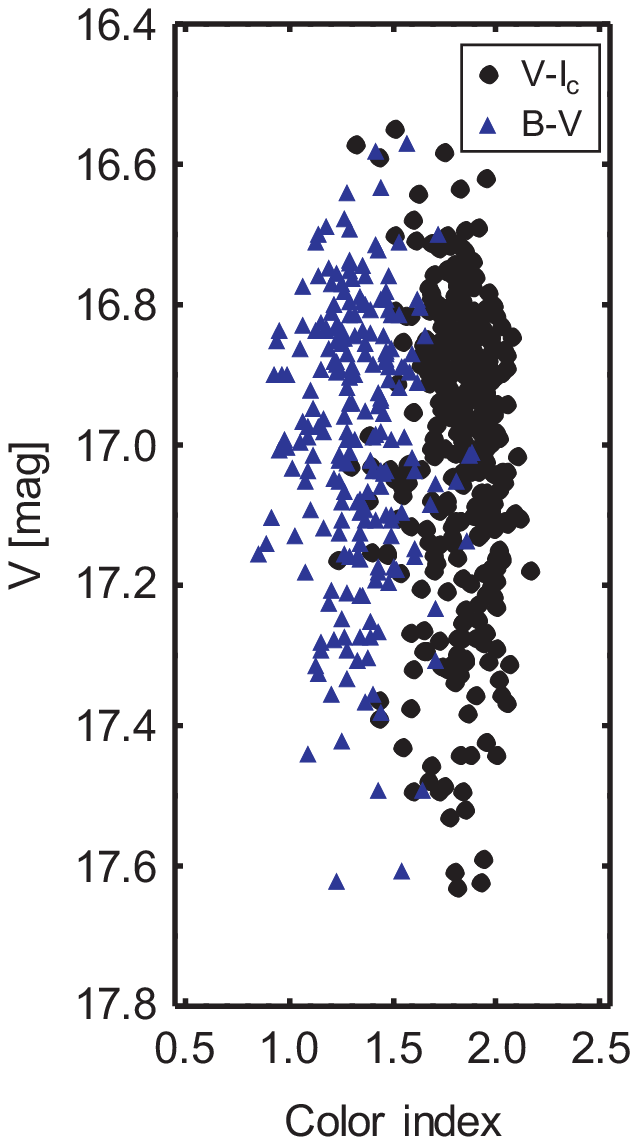}
		\caption{$BVR_{c}I_{c}$ light curves (left) and color$-$magnitude diagrams (right) of 2MASS J21383255+5730161.}\label{Fig:curves4}
	\end{center}
\end{figure}

\begin{figure}[h!]
	\begin{center}
		\includegraphics[width=11cm]{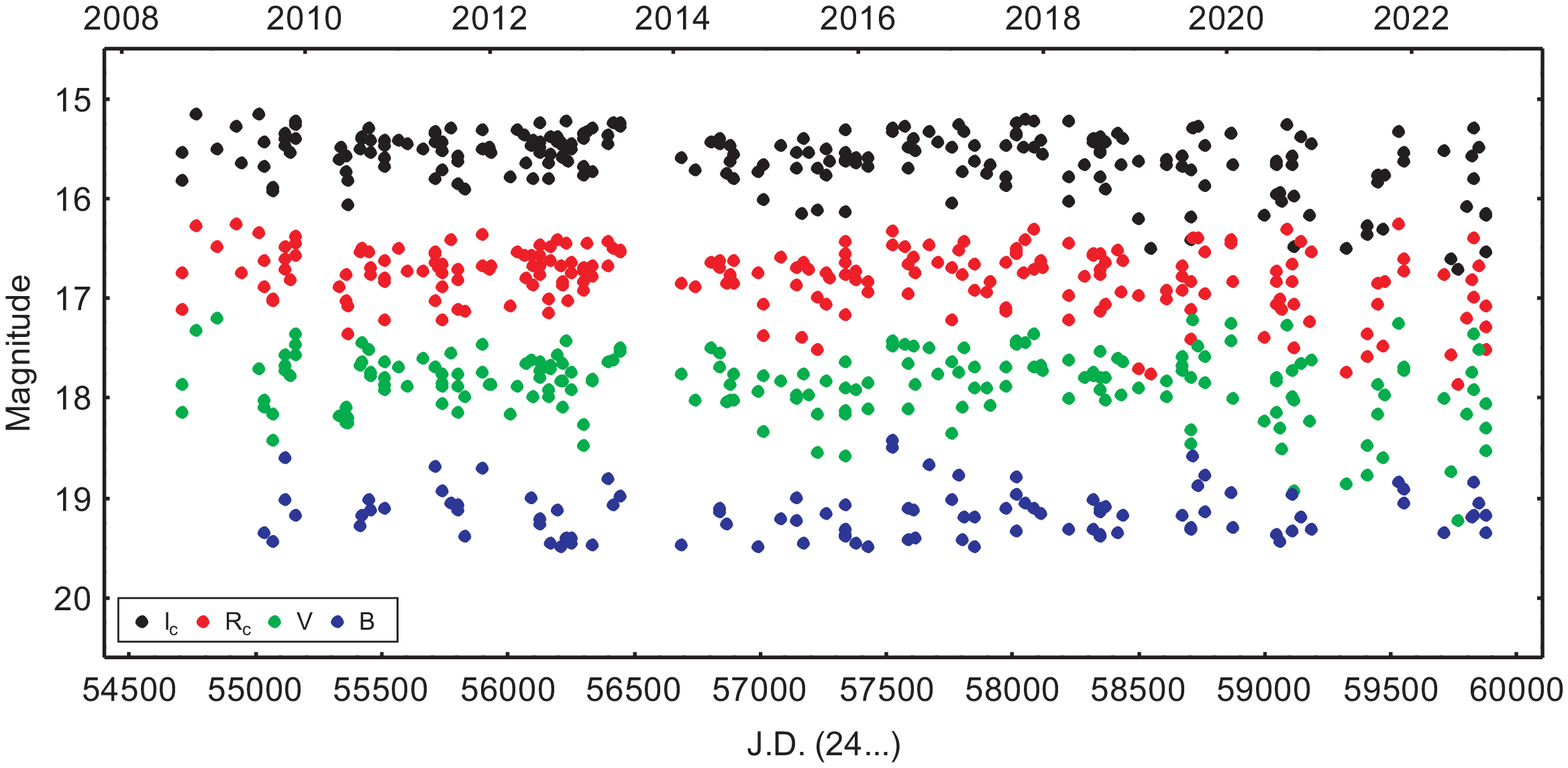}
		\includegraphics[width=3cm]{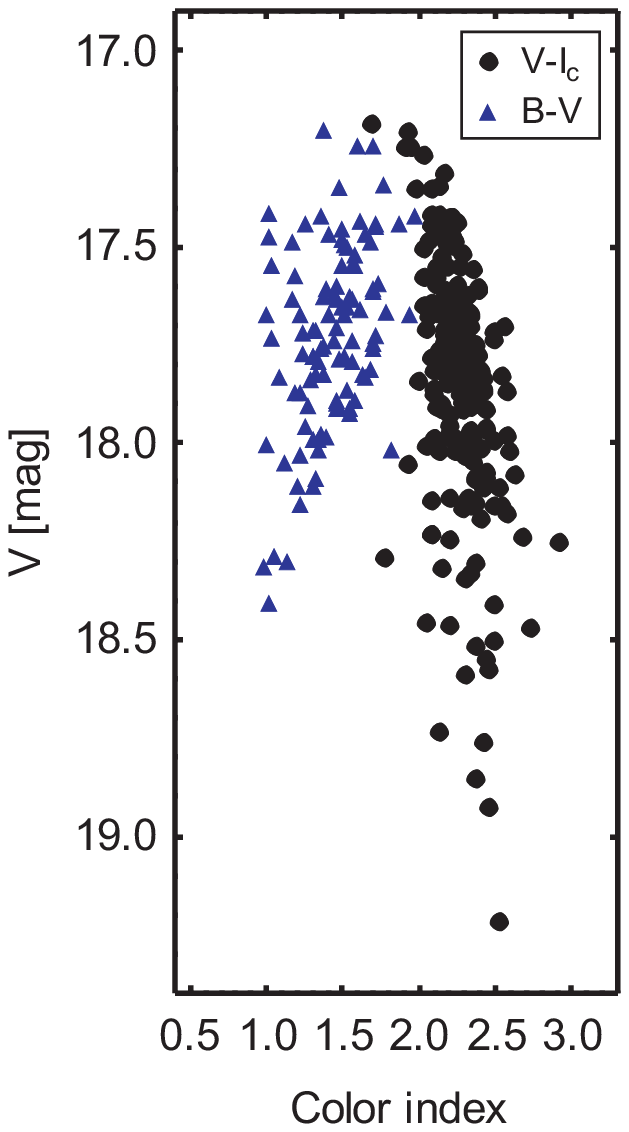}
		\caption{$BVR_{c}I_{c}$ light curves (left) and color$-$magnitude diagrams (right) of 2MASS J21393480+5723277.}\label{Fig:curves5}
	\end{center}
\end{figure}

\begin{figure}[h!]
	\begin{center}
		\includegraphics[width=11cm]{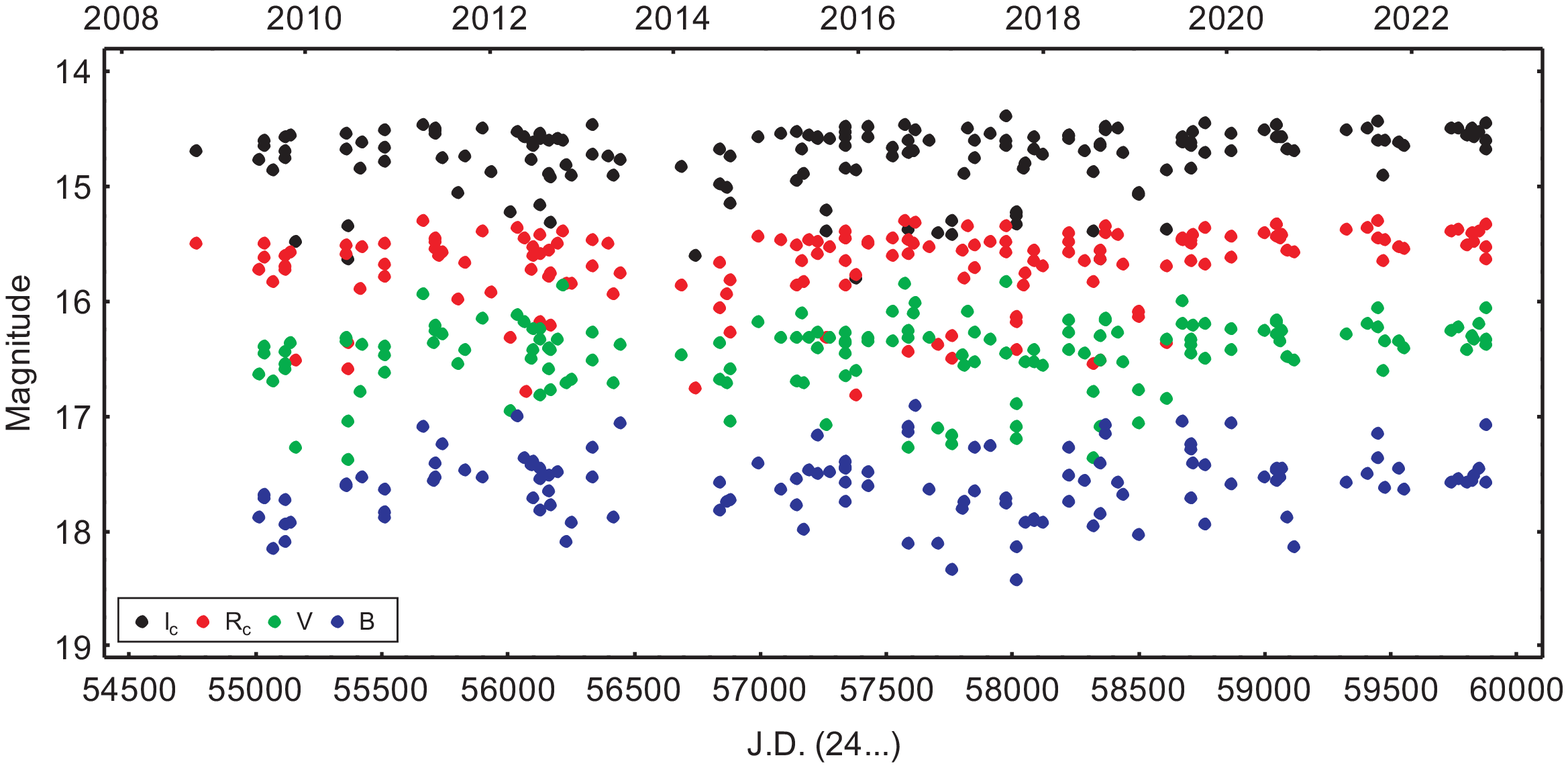}
    	\includegraphics[width=3cm]{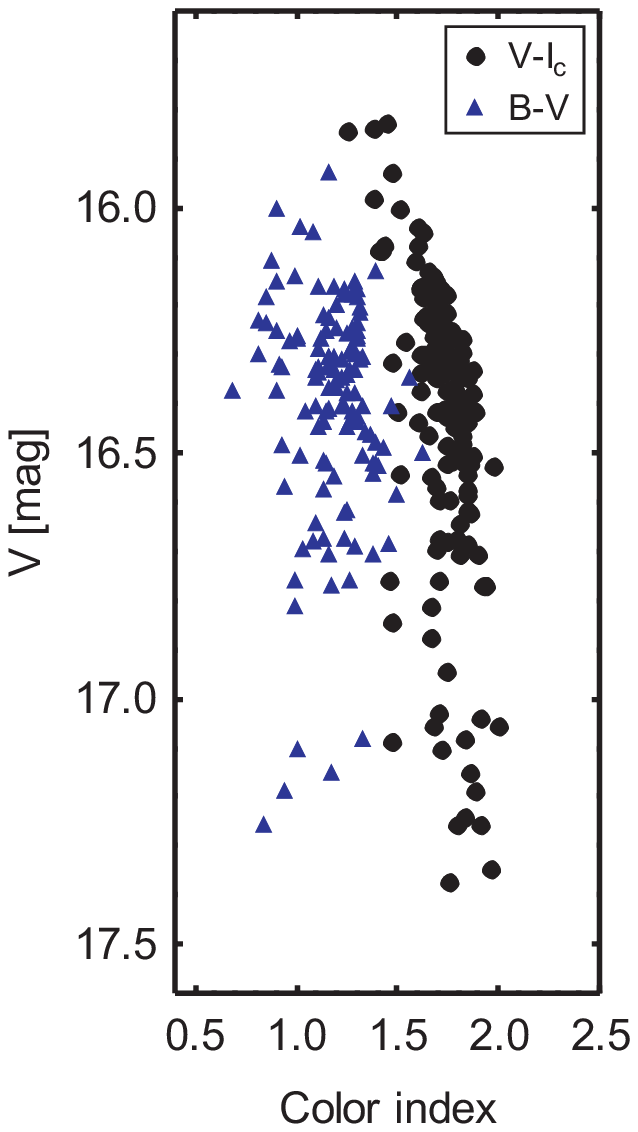}
		\caption{$BVR_{c}I_{c}$ light curves (left) and color$-$magnitude diagrams (right) of 2MASS J21403134+5733417.}\label{Fig:curves7}
	\end{center}
\end{figure}

The registered amplitudes of the light variations of 2MASS J21383255+5730161, 2MASS J21393480+5723277 and 2MASS J21403134+5733417 are given in Table~\ref{Tab:amplitudes}.
As can be seen, the brightness of the stars varies in wide ranges.
The large amplitude dips in brightness are an indication of UXor type variability.
In this case, it can be assumed that at least one of the reasons for the photometric variability of 2MASS J21383255+5730161, 2MASS J21393480+5723277 and 2MASS J21403134+5733417 is the variable extinction.

During our photometric monitoring, we registered a change in the color of 2MASS J21383255+5730161, 2MASS J21393480+5723277 and 2MASS J21403134+5733417 at their brightness minima (Fig.~\ref{Fig:curves4}, Fig.~\ref{Fig:curves5}, Fig.~\ref{Fig:curves7}).
It is seen, that initially these stars become redder as they fade, but during the minima in their brightness, the blueing effect is registered.
It can be assumed that at least one of the causes for the registered dips in the photometric behavior of the stars is the variation in the density of dust and clouds in orbit around the star, which crosses the line of sight and obscured the objects.
We did not detect any periodicity in the variability of 2MASS J21383255+5730161, 2MASS J21393480+5723277 and 2MASS J21403134+5733417.

\section{Conclusion remarks}
\label{sect:conclusion remarks}

We investigated the $BV(RI)_{c}$ photometric variability of eight young stars in the young open stellar cluster Trumpler 37.
All stars from our study manifest typical PMS stars' photometric variability.
The main results of our study are as follows:
(i) 2MASS J21365072+5731106 (LkH$\alpha$ 349) exhibits photometric variability with amplitude under 0.2 mag in all-optical passbands, typical for weak-line TTSs and HAEBESs.
For more than 50 years, the star's brightness maintains the same level.
(ii) 2MASS J21381703+5739265 and 2MASS J21403574+5734550 manifest strong irregular variability, typical for TTSs.
(iii) 2MASS J21382596+5734093 and 2MASS J21393612+5731289 show rapid brightness variation, typically for TTSs, probably caused by rotation modulation of cool spot(s) on the stellar surface.
We confirmed the periodicity of 0.9030$^d$ for 2MASS J21382596+5734093 and we registered a periodicity of 1.4919$^d$ for 2MASS J21393612+5731289.
(iv) the variability of 2MASS J21383255+5730161, 2MASS J21393480+5723277 and 2MASS J21403134+5733417 is characterized by multiple brightness dips, typical for UXor type stars.
We plan to continue our photometric multicolor monitoring of the investigated stars during the next years.

\begin{acknowledgements}
This research has made use of NASA's Astrophysics Data System Abstract Service.
The authors thank the Director of Skinakas Observatory Prof. I. Papamastorakis and Prof. I. Papadakis for the award of telescope time.
We acknowledge with thanks the variable star observations from the AAVSO International Database contributed by observers worldwide and used in this research.
This work was partly supported by the research fund of the University of Shumen, Bulgaria.
We thank the anonymous referee for carefully reading the text and for the useful suggestions and comments that helped to improve the paper.
\end{acknowledgements}

\appendix                  


\begin{thebibliography}{36}
	
	\bibitem[2011]{bare11} Barentsen G., Vink J. S., Drew J. E., Greimel R., Wright N. J., Drake J. J., Martin E. L., Valdivielso L., Corradi R. L. M., 2011, MNRAS, 415, 103
	
	\bibitem[1990]{bibo90} Bibo E. A., Th\'{e} P. S., 1990, A\&A, 236, 155
	
	\bibitem[2020]{chen20} Chen X., Wang S., Deng L., de Grijs R., Yang M., Tian H., 2020, ApJ Suppl. Ser., 249, 18
	
	\bibitem[1979]{cohe79} Cohen M., Kuhi L. V., 1979, ApJ Suppl. Ser., 41, 743
	
	\bibitem[2002]{cont02} Contreras M. E., Sicilia-Aguilar A., Muzerolle J., Calvet N., Berling P., Hartmann L., 2002, AJ, 124, 1585
	
	\bibitem[1989]{cram89} Cram L. E., Kuhi L. V., Jordan S., Thomas R., Goldberg L., Pecker J.-C., 1989, Monograph Series on Nonthermal Phenomena in Stellar Atmospheres: FGK stars and T Tauri stars (NASA SP-502)
	
	\bibitem[1968]{diba68} Dibai \'{E}., Esipov V. F., 1968, Soviet Astronomy, 12, 448
	
	\bibitem[2003]{dull03} Dullemond C. P., van den Ancker M. E., Acke B., van Boekel R., 2003, ApJ, 594, L47
	
	\bibitem[1995]{fern95} Fernandez M., 1995, A\&A Suppl., 113, 473
	
	\bibitem[1996]{fern96} Fernandez M., Eiroa C., 1996, A\&A, 310, 143
	
	\bibitem[1991]{grin91} Grinin V. P., Kiselev N. N., Minikulov N. Kh., Chernova G. P., Voshchinnikov N. V., 1991, Ap\&SS, 186, 283
	
	\bibitem[1972]{herb72} Herbig G. H., Kameswara Rao N., 1972, ApJ, 174, 401
	
	\bibitem[1994]{herb94} Herbst W., Herbst D. K., Grossman E. J., Weinstein D., 1994, AJ, 108, 1906
	
	\bibitem[1995]{hess95} Hessman F. V., Beckwith S. V. W., Bender R., Eisl\"{o}ffel J., G\"{o}tz W., Guenther E., 1995, A\&A, 299, 464
	
	\bibitem[2015]{ibry15} Ibryamov S. I., Semkov E. H., Peneva S. P., 2015, PASA, 32, e021
	
	\bibitem[2020]{ibry20} Ibryamov S., Semkov E., 2020, Bulgarian Astronomical Journal, 32, 96
	
	\bibitem[1945]{joy45} Joy A. H., 1945, ApJ, 102, 168
	
	\bibitem[1997]{kaza97} Kazarovets E. V., Samus N. N., 1997, IBVS, 4471, 1
	
	\bibitem[2005]{lenz05} Lenz P., Breger M., 2005, Communications in Asteroseismology, 146, 53
	
	\bibitem[1975]{lore75} Loren R. B., Peters W. L., Vanden Bout P. A., 1975, ApJ, 195, 75
	
	\bibitem[1990]{mars90} Marschall L. A., Comins N. F., Karshner G. B., 1990, AJ, 99, 1536
	
	\bibitem[1999]{mena99} M\'{e}nard F., Bertout, B., 1999, in NATO Advanced Science Institutes (ASI) Series C, 540, ed. C. J. Lada \& N. D. Kylafis, 341
	
	\bibitem[2019]{meng19} Meng H. Y. A., Rieke G. H., Kim J. S., Sicilia-Aguilar A., Cross N. J. G., Esplin T., Rebull L. M., Hodapp K. W., 2019, ApJ, 878, 29
	
	\bibitem[2022]{muta22} Mutafov A., Semkov E., Peneva S., Ibryamov S., 2022, Bulgarian Astronomical Journal, 36, 3
	
	\bibitem[1989]{naka89} Nakano M., Tomita Y., Ohtani H., Ogura K., Sofue Y., 1989, PASJ, 41, 1073
	
	\bibitem[1995]{pate95} Patel N. A., Goldsmith P. F., Snell R. L., Hezel T., Xie T., 1995, ApJ, 447, 721
	
	\bibitem[2003]{petr03} Petrov P. P., 2003, Ap, 46, 506
	
	\bibitem[1996]{schw96} Schwarzenberg-Czerny A., 1996, ApJ, 460, L107
	
	\bibitem[2012]{semk12} Semkov E. H., Peneva S. P., 2012, Ap\&SS, 338, 95
	
	\bibitem[2015]{semk15} Semkov E. H., Ibryamov S. I., Peneva S. P., Milanov T. R., Stoyanov K. A., Stateva I. K., Kjurkchieva D. P., Dimitrov D. P., Radeva V. S., 2015, PASA, 32, e011	
	
	\bibitem[2005]{sici05} Sicilia-Aguilar A., Hartmann L. W., Her\'{a}ndez J., Brice\~{n}o C., Calvet N., 2005, AJ, 130, 188
	
	\bibitem[2006]{sici06} Sicilia-Aguilar A., Hartmann L. W., F\"{u}r\'{e}sz G., Henning T., Dullemond C., Brandner W., 2006, AJ, 132, 2135
	
	\bibitem[2010]{sici10} Sicilia-Aguilar A., Henning T., Hartmann L. W., 2010, ApJ, 710, 597
	
	\bibitem[1968]{simo68} Simonson S. C. III., 1968, ApJ, 154, 923
	
	\bibitem[1989]{vosh89} Voshchinnikov N. V., 1989, Astrofizika, 30, 509
	
	\bibitem[1986]{zait86} Zaitseva G. V., 1986, Ap, 25, 626

\end{thebibliography}
\end{document}